\documentclass[12pt,preprint]{aastex}

\shorttitle{New Class of Pulsating Subdwarf B Stars}
\shortauthors{Green et al.}

\newcommand{\teff}{$T_{\rm eff}$}
\newcommand{\logg}{log~{\it g}}

\begin{document}

\title{Discovery of A New Class of Pulsating Stars: \\
  Gravity-mode Pulsators among Subdwarf B Stars}

\author{E. M. Green\altaffilmark{1}, 
G. Fontaine\altaffilmark{2}, M. D. Reed\altaffilmark{3}, K. Callerame\altaffilmark{1}, I. R. Seitenzahl\altaffilmark{1}, 
B. A. White\altaffilmark{1}, E. A. Hyde\altaffilmark{1}, R. \O stensen\altaffilmark{4}, O. Cordes\altaffilmark{5}, 
P. Brassard\altaffilmark{2}, S. Falter\altaffilmark{6}, E. J. Jeffery\altaffilmark{7}, S. Dreizler\altaffilmark{8},
S. L. Schuh\altaffilmark{8}, M. Giovanni\altaffilmark{1}, H. Edelmann\altaffilmark{6}, J. Rigby\altaffilmark{1},
and A. Bronowska\altaffilmark{4} }

\email{egreen@as.arizona.edu}

\altaffiltext{1}{Steward Observatory, University of Arizona, Tucson, AZ 85721; 
egreen@as.arizona.edu; callerame@cox.net; irs@uchicago.edu;
bwhite@as.arizona.edu; elainahyde@yahoo.com; jrigby@as.arizona.edu}

\altaffiltext{2}{D\'epartement de Physique, Universit\'e de Montr\'eal, CP6128, Station 
Centreville, Montr\'eal, QC H3C 3J7, Canada;
fontaine@astro.umontreal.ca; brassard@astro.umontreal.ca}

\altaffiltext{3}{Physics, Astronomy, \& Material Science, SW Missouri State University, 
Springfield, MO 65804; mreed@sdbv.smsu.edu}

\altaffiltext{4}{Isaac Newton Group of Telescopes, 37800 Santa Cruz de La Palma,
Canary Islands, Spain; roy@ing.iac.es; bronka@tiger.chem.uw.edu.pl}

\altaffiltext{5}{Sternwarte of the University Bonn, Auf dem Huegel 71, D-53123 Bonn, 
Germany; ocordes@astro.uni-bonn.de}

\altaffiltext{6}{Dr.-Remeis-Sternwarte University Erlangen-N\"urnberg, Sternwartstr.~7,
D-96049 Bamberg, Germany; falter@sternwarte.uni-erlangen.de}

\altaffiltext{7}{Brigham Young University, Provo, UT 84602; ejj9@email.byu.edu}

\altaffiltext{8}{Institut f\"ur Astronomie und Astrophysik, Universit\"{a}t T\"ubingen, 
Sand 1, D-72076 T\"ubingen, Germany;  
dreizler@astro.uni-tuebingen.de; schuh@astro.uni-tuebingen.de}

\begin{abstract}

During the course of an ongoing CCD monitoring program to investigate
low-level light variations in subdwarf B (sdB) stars, we have
serendipitously discovered a new class of low amplitude, multimode sdB
pulsators with periods of the order of an hour.  These periods are
more than a factor of ten longer than those of previously known
multimode sdB pulsators (EC\,14026 stars), implying that they are due
to gravity modes rather than pressure modes.  The longer period
pulsators are found only among cooler sdB stars, where they are
surprisingly common.  The iron opacity instability that drives the
short period EC\,14026 stars is effective only in hot sdB's, leaving
the driving mechanism for the deeper gravity modes in cool sdB's
currently unknown.  We present the first observational results for
our newly identified sdB variables, and discuss possible
implications.

\end{abstract}



\section{Introduction}

Subdwarf B (sdB) stars are evolved hot stars (22,000 to 40,000\,K)
commonly found in the disk of our Galaxy (e.g.\ Saffer et al.\ 1994).
They are widely believed to be core He-burning stars of $\sim
0.5\,M_{\sun}$ on the extended horizontal branch (EHB) (Heber 1986;
Saffer et al.\ 1994), having lost nearly all of their envelopes near
the first red giant tip.  The existence of multimode pulsators among
sdB stars presents a powerful opportunity for further investigation of
their structure using asteroseismological techniques.  In addition to
clarifying the uncertain evolutionary histories of the sdB stars
themselves, the results will be relevant to the later evolution of the
vast majority of normal stars.  Buried under their huge envelopes, all
low mass He-burning red giants have cores that are basically
identical to sdB stars.  It is our good fortune that the latter have
shed all but a few wisps of their obscuring envelopes, and that some
appear to pulsate with exactly the right complexity
required for asteroseismology.

The first pulsating sdB stars were discovered only recently (Kilkenny
et al.\ 1997).  Commonly called EC\,14026 stars, after the
prototype\footnote{EC\,14026 stars are now officially V361 Hya
stars.}, they are multimode pulsators with typical periods of
100--250~s, and a total range of about 80--600~s.  Independently and
nearly simultaneously, their existence was predicted by Charpinet et
al.\ (1996, 1997), due to the presence of low-order and low-degree
radial and nonradial pressure modes (p-modes) excited by an iron
opacity $\kappa$~mechanism in the thin diffusion-dominated envelopes.
Additional EC\,14026 stars were soon found by teams at the South
African Astronomical Observatory (Kilkenny 2002, and references
therein), the University of Montreal (Bill\`eres et al.\ 2002), and
the Nordic Optical Telescope (NOT) (Silvotti et al.\ 2002, and
references therein), bringing the current total to about 30.  The
justifiable excitement over these stars (Koen et al.\ 1998a; Fontaine
et al.\ 1998; O'Donoghue et al.\ 1999; Charpinet, Fontaine, \&
Brassard 2001; Kilkenny 2002) stems from the fact that stellar and
atmospheric models for sdB stars appear to be sufficiently realistic
for nonadiabatic calculations to be extremely successful in explaining
their properties (Charpinet et al.\ 2001).  In agreement with the
theoretical results, EC\,14026 variables are found primarily among the
hotter sdB stars, clustering around \teff\ $\sim$ 33,500\,K and \logg\
$\sim$ 5.8, with only a few low gravity examples at somewhat cooler
temperatures.\footnote{PG\,1605+072 (Koen et al.\ 1998b); Feige\,48
(Koen et al.\ 1998c); HS\,2201+2610 (\O stensen et al.\ 2001a);
HS\,0702+6043 (Dreizler et al.\ 2001)} The pulsation amplitudes are
generally less than a few hundredths of a magnitude.

In this letter, we introduce another chapter in sdB asteroseismology:
the discovery of a new class of multimode pulsators.  The newly
identified stars have apparent periods at least a factor of ten longer
than those observed in EC\,14026 stars, implying that they must be due
to gravity modes (g-modes).  The longer periods present both
advantages and disadvantages for further observations.  The most
important gain may be the greatly increased range and number of
suitable EHB candidates for asteroseismology.  The following sections
describe the observations and apparent trends identified in the data,
discuss their implications for the unknown excitation mechanism, and
summarize our conclusions.  Subsequent papers will examine individual
pulsators in more detail.

\section{Observational Characteristics}

The prototype long period sdB variable, PG\,1716+426, is an
unremarkable faint blue star less than 1$\degr$\ south of M92 in
Hercules ($V\, \simeq$\, 13.97, Wesemael et al.\ 1992).  Its
variability was first detected in 1999, though its nature wasn't fully
appreciated for nearly two years.  The discovery light curve (shown at
the top of Figure~1) was taken amidst thin cirrus at the Steward
2.3\,m telescope on Kitt Peak, during initial tests of a pilot program
to monitor sdB binaries with large radial velocity variations.  We
were searching for eclipses by white dwarfs, ellipsoidal effects, and
reflection effects, all of which have amplitudes less than a few
hundredths of a magnitude and timescales longer than 250~s.  Except
where noted, all data for this paper were obtained at the Arizona
2.3\,m and 1.6\,m telescopes on Kitt Peak and Mt.\ Bigelow
using conventional 2K\,CCD's binned 3x3
(0.45\arcsec/pixel) to reduce the overhead time to 24~s.  Depending on
the target, integration times of 15--60~s were chosen to maximize the
counts and minimize photometric errors.  Differential magnitudes were
derived from aperture photometry of the sdB relative to reference
stars of comparable magnitude within each frame.

Six additional light curves for PG\,1716+426, 2--4 hours each during
April--June 2000, showed persistently irregular variations in both
amplitude and phase, that were not present in any of several nearby
reference stars.  Both low resolution (Saffer et al.\ 1994) and very
high S/N 1\AA\ resolution MMT spectra (Green et al., in preparation),
and multicolor photometry (Wesemael et al.\ 1992) proved that
PG\,1716+426 is a normal, moderately cool sdB star, but the timescales
seemed much too long for an EC\,14026 star.  Eventually we detected
similar variations in a second sdB, PG\,0101+039, although at lower
amplitude.  We were allocated 18 nights at Mt.\ Bigelow to monitor
PG\,0101+039, but were able to observe for only one night due to
unseasonable bad weather.  We finally acquired 81 hours of CCD
photometry for PG\,1716+426 during 14 nights in April--June 2001: 12
clear nights at the Mt.\ Bigelow 1.6\,m telescope, several adjoining
hours at the end of a partially cloudy night on the Calar Alto 2.2\,m,
a separate night on the NOT, and another with the Fick Observatory
0.6\,m telescope.  At Mt.\ Bigelow and at Fick, all data were taken
with an $R$ filter.  The NOT observations cycled rapidly between $V$,
$R$, and $I$ with ALFOSC (\O stensen et al.\ 2001b), while the BUSCA
camera at Calar Alto allowed simultaneous photometry in four wide
passbands from $U$ to $I$ (Falter et al.\ 2003).  A first analysis of
all the $R$ data found at least 3--5 pulsation modes with periods
between 0.8 and 1.4 hours.  Further details will be presented by Reed
et al.\ (in preparation).

Once pulsations were confirmed in PG\,1716+426, we broadened the
1.6\,m monitoring program to include a nearly random sample of bright
sdB stars ($V\, \le\, 14$).  We have now identified 20 stars
(including PG\,1716+426) that appear to be multimode pulsators with
periods of the order of an hour, from a total sample of 72.  Each star
was monitored for 2 to 7 hours at a time, and most were observed on
multiple nights to allow for the possibility of occasional
cancellation of modes.  (PG\,1716+426 exhibited a nearly flat light
curve for nearly four hours once, in 14 nights of observation.)
The peak-to-peak amplitudes in the light curves are comparable with
those of most EC\,14026 stars.  The strongest new pulsators clearly
show the characteristic beating effects of multiple modes with close
periods (Fig.~1).  However, about half of our variables pulsate
so weakly that relative light variations are detectable only because
of the stability, precision, and good sky coverage of modern CCD's.
We classify the weakest pulsators as bonafide members of the class
only when light variations are clearly detected on more than one
night, but even in their ``low'' states, the light curves are usually
not as flat as those of non-pulsators.  Daily aliasing is obviously a
major problem, since fewer than 8 to 10 cycles per night can be
observed from a single site.  Extensive, well-coordinated multi-site
campaigns or space observations over fairly long time baselines are
required for good mode characterization.

The only selection biases arise from a preference for brighter
targets, mostly from the PG catalog (Green, Schmidt, \& Liebert 1986),
and our avoidance of sdB stars with composite spectra (for better
temperature and gravity determinations).  Otherwise, we selected stars
having the best sky positions at the time of observation.  We
therefore expect our sample to be representative of all sdB stars for
properties such as temperature, gravity, presence of close
companions, etc.  For example, PG\,1716+426\footnote{We derive an
orbital period of 1.77739$^{\rm d}$ for PG\,1716+426; Morales-Rueda et
al.\ (2002) found 1.77732$^{\rm d}$.} and the next few members of the
class were all found in single-lined spectroscopic binaries with
orbital periods of less than a few days, but this was due only to our
pre-discovery sample selection.  Now that we have obtained light
curves for most of a representative subsample of 52 sdB stars for
which we have precise radial velocities from multiple spectra, we find
no binary correlation.  Specifically, 29 of the 52 sdB stars (56\%)
are short period binaries, having the same apparent distribution in
temperature and gravity as the rest of the sample.  13 of the 52 are
also long period pulsators.  If the pulsations are independent of
close binary status, we expect 7.25~$\pm$~2.4 binary pulsators.  The
observed number is 7, indicating that the fraction of short period
binaries among PG\,1716-like stars is no different from that of sdB
stars in general.

The present data suggest an upper limit of 90--120 minutes for the
major periods of all the new pulsators.  We have not found any sdB
pulsators with obvious periods as long as the 7.8 hour sawtooth
modulation detected by Orosz \& Wade (1999) in KPD\,0422+5421,
although we did find three low amplitude variables with periods of
several hours or more.  Two are sdB stars exhibiting apparent
reflection effects, having appropriately phased sinusoidal light
curves the same length as their orbital periods ($\sim$ 6$^{\rm h}$
and 12$^{\rm h}$).  The third was a reference star with an early F
spectrum and a 6$^{\rm h}$ period, a likely $\delta$~Scuti star.

A general characteristic of nonradial pulsators is the expected
amplitude dependence of the pulsations as a function of wavelength,
although the largest effects are in the ultraviolet (e.g.\ Robinson et
al.\ 1995; Brassard, Fontaine, \& Wesemael 1995).  Multicolor
amplitudes have been reported for two short period (p-mode) pulsators.
Koen (1998) observed somewhat larger $U$ amplitudes in KPD\,2109+4401,
but no significant trends in $BVR$.  Falter et al.\ (2003) found
similar behavior in PG\,1605+072.  Following Brassard et al.\ (1995),
we computed the wavelength dependence of the amplitude of a g-mode in
a 30,000\,K, \logg\ = 5.5 sdB model.  We find that the theoretical
curve is rather flat in the red and increases somewhat toward the
blue: predicted amplitude ratios are 1.09 for $B/V$, 1.38 for $U/V$.
Our observations appear mostly consistent.  The NOT data for
PG\,1716+426 show no obvious systematic trends in $VRI$, and
(non-simultaneous) 2.3\,m $BVR$ amplitudes are fairly comparable.  At
first glance, the largest amplitude in the short and somewhat noisier
BUSCA data appears to be in the $I$ filter, but closer inspection
reveals that the amplitudes cannot be disentangled from surprisingly
large slopes in the baselines (mainly in the wide $U$ and $I$
filters).  Subsequent experiments with simultaneous $B$ and $R$
photometry in Arizona demonstrated that the baseline trends are almost
certainly due to differential extinction between the sdB and the much
redder comparison stars.  With hour long timescales, there are serious
problems at bluer wavelengths or with wider bandpasses, because the
effect of differential extinction is comparable to the pulsational
amplitudes.  At Mt.\ Bigelow and Kitt Peak, we generally see
significant differential extinction in $B$, a less noticeable effect
in $V$, and no detectable effect in the (preferred) $R$ filter ($I$ is
usually avoided due to fringing).

Figure~2 compares the positions of the long period and short period
sdB pulsators in the \logg\ vs \teff\ diagram.  The 13 newly
identified pulsators (red and green error bars) for which we derived
homogeneous temperatures and gravities are all cooler than 30,000\,K,
in marked contrast with the hotter EC\,14026 stars (blue filled
circles).  We lack spectra for seven (unplotted) new pulsators, but
six of those with temperatures from other sources also appear to be
relatively cool (Heber 1986; Wesemael et al.\ 1992; Saffer et al.\
1994; Morales-Rueda et al.\ 2002).  At first glance, there seems to be
little overlap in temperature and gravity between the long and short
period pulsators, and no long period variations have been found in any
EC\,14026 stars\footnote{This is a weak constraint, since only a few
EC\,14026 stars are bright enough for our survey, and several of these
have such complicated light curves that it would be very difficult to
detect additional long period components, e.g.\ PG\,1605+072 (extreme
amplitudes), PG\,1336-018 (eclipsing sdB + dM), or KPD\,1930+2752 (sdB
+ wd with ellipsoidal variations).}.  However, we note that the
inhomogeneous atmospheric parameters for the latter (Charpinet 2001)
are not necessarily on the same scale as our data points.  Our
temperatures are systematically somewhat hotter than those derived
by some other investigators for \teff\ $<$ 32,000\,K, and our
gravities consequently slightly higher.  (We are continuing to pursue
the longstanding disagreement between various atmospheric codes for
sdB stars (Wesemael et al.\ 1997), but it is beyond the scope of 
this paper.)  Thus, it is probably premature to speculate about
the (small) degree of separation or overlap between the two types of
pulsators in Figure~2.  The important point is that long period sdB
pulsators clearly populate a separate temperature and gravity regime
than do short period pulsators, in addition to their very different
timescales.

The fraction of sdB stars exhibiting long period light variations is
surprisingly high, in further contrast with the EC\,14026 stars:
roughly 75\% of sdB stars cooler than 30,000\,K, or 25--30\% of all
sdB stars.

\section{Discussion}

Observed periods of the order of an hour require the newly
discovered sdB variables to be gravity mode pulsators.  Preliminary
calculations show that high radial order g-modes are needed to produce
the range of periods observed in PG\,1716+426.  The $\kappa$~mechanism
that has been so successful for understanding p-mode pulsations in
EC\,14026 stars (Charpinet et al.\ 1997; Brassard et al.\ 2001) does
not appear to be capable of driving such deep pulsational modes.
The g-mode driving mechanism is therefore unknown.

The limited temperature range of the long period variables in Figure~2
implies the pulsation mechanism is intrinsic to the star, rather than
an external driver such as tidal excitation (Fontaine et al.\ 2003).
The lower temperatures result from increased hydrogen envelope masses,
up to 0.003--0.004\,$M_{\sun}$ in the coolest sdB stars.  However, it
is not clear how such apparently negligible envelopes could cause 
longer period pulsations.

Canonical EHB evolutionary sequences (e.g.\ Dorman's tracks, 
from Charpinet et al.\ (2000), shown in Figure~2) indicate that a
star initially expands away from the zero age EHB, and then begins
contracting toward higher temperatures as helium is consumed in the
core.  Following central helium exhaustion (where the slope of the
tracks changes sign for the second time, very abruptly), helium shell
burning produces a feeble attempt at further expansion, which is
quickly succeeded by rapid contraction to the white dwarf stage.  Our
data suggest that larger amplitude g-mode pulsators (in red) 
occur preferentially (only?)\ toward the end of core helium burning
(after correcting for a small apparent offset by shifting the
theoretical tracks down until their starting points form a lower
envelope to all the cooler stars).  If strong pulsators do not exist
near the zero age EHB, it is tempting to think that they might be
driven by a slowly reviving hydrogen burning shell.  However, in
standard EHB models, the miniscule envelope does not allow the
hydrogen shell to ``turn on'' until after the star has contracted to a
very hot white dwarf.  Thus, if the hydrogen envelope is really of key
importance in the driving mechanism, its role could be elucidated only
with nonstandard models of sdB stars.  It would be interesting to know
if similar g-mode pulsations can occur in sdB stars whose spectra show
little if any hydrogen: the He-sdB, or sdB:He4 stars (Jeffery et al.\
1997, and references therein), if any exist with comparable
temperatures and gravities.

P.\ Moskalik (private communication) pointed out
that the distribution of short- and long-period sdB pulsators in
the \logg\ vs \teff\ diagram resembles the situation in main sequence
B stars.  In both cases, the hotter objects are low-order p-mode
pulsators, while the cooler stars are high-order g-mode pulsators.
Dziembowski, Moskalik, \& Pamyatnykh (1993) found that the same
mechanism could drive both types of pulsation on the main sequence.
Perhaps a variant of Charpinet et al.'s (1997) $\kappa$~mechanism
could be found to work in the cooler sdB stars, though this remains to
be seen.

\section{Summary and Conclusions}

We have so far identified 20 subdwarf B stars that appear to be
multimode pulsators on timescales of about an hour, more than a
factor of ten longer than typical periods of the recently discovered
EC\,14026 pulsators.  Observed peak-to-peak amplitudes are $\la$~0.05
magnitudes.  Due to the longer timescales, only the prototype,
PG\,1716+426, has been sufficiently well observed to
detect individual modes.  Our first attempt, with 81 hours monitoring
spread over 2 months, indicates the presence of at least 3--5 probable
modes with periods between 0.8 and 1.4 hours.  All long period sdB
pulsators with known, spectroscopically determined temperatures are
relatively cool (T$_{\rm eff}$ $<$ 30,000\,K), in contrast with
the hotter EC\,14026 stars.  PG\,1716 stars are also about six times
more common.

Many of the newly identified variables are easily observed with small
telescopes capable of precise differential photometry.  Fast frame
transfer CCD's are not required; conventional CCD's may have an
advantage if their larger field of view permits a better selection of
reference stars.  Good mode characterization will require photometry
over a much larger fraction of a day than can be obtained from a
single site, in addition to long time baselines.  With ground-based
campaigns, differential extinction is a much worse problem for long
period sdB pulsators than for the short period ones, but it can be
minimized by using an R filter.  PG\,1716 stars are excellent
candidates for observation from space, to eliminate both aliasing and
atmospheric problems.

The long periods found in PG\,1716+426, and strongly indicated in the
light curves of nineteen other cool sdB variables, imply the detection
of high radial order g-modes.  The driving mechanism for g-modes in
sdB stars is currently unknown.  The distribution of the stars in the
\logg\ -- \teff\ plane suggests that their longer pulsations might be
related to thicker hydrogen envelopes, but this is not easily
understood from standard EHB models.  Alternatively, some variant of
the EC\,14026 iron opacity mechanism might also be at work in cooler
sdB stars, analogous to the situation in main sequence B stars
(Dziembowski et al.\ 1993).

The relatively long pulsation timescales, the distinctly different
temperature range, and the inability of the existing iron opacity
mechanism to drive g-modes, all support the identification of the long
period sdB stars as a separate class of pulsating variables, one with
great potential to expand our understanding of evolved stars.

\acknowledgments

This investigation was made possible by NSF grants AST-9731655 and
AST-0098699.  We thank the Steward TAC for their generous time
allocations, and P.\ Strittmatter for his continuing support of
undergraduate research.  We are grateful to the SO staff, especially
E.\ Christensen, B.\ Wood, and G.\ Rosenbaum, for their patience with
our lengthy calibrations, and all their other assistance.  EMG thanks
Bill Peters for many tangible and intangible contributions too
numerous to mention.

\clearpage

\figcaption{The discovery light curve for PG\,1716+426 (top, observed
with a V filter), followed by typical lightcurves for the largest
amplitude long period pulsators: PG\,1716+426~(R filter),
PG\,0850+170~(R), PG\,1338+481~(B), PG\,1627+017~(R), PB\,5450~(R),
PG\,1739+489~(V).  The time between peaks varies from about 35 to
nearly 120 minutes. }

\figcaption{The \logg\ vs \teff\ diagram for various pulsating sdB
stars superposed on Dorman's EHB evolutionary tracks (light dotted
lines) for models with a range of H envelope masses.  The thick
2$\sigma$ error bars represent the long period pulsators for which we
have derived precise, homogeneous atmospheric parameters using high
S/N MMT spectra.  The stronger pulsators are plotted in red, and
weaker pulsators in green.  Thin black error bars show the
non-pulsating sdB stars in our sample.  The filled blue circles are
the known short period EC\,14026 pulsators, from a variety of sources
(Charpinet 2001).  The thicker dotted line represents the zero-age
He-burning main sequence (i.e.\ the zero hydrogen envelope limit to
the EHB).}

%%%UCP%%%
%\newpage
%\plotone{f1.eps}
%\newpage
%\plotone{f2.eps}

%\notetoeditor{Figure 1 is in landscape mode.}

%\notetoeditor{Figure 2 is a color plot.}

%\vfill

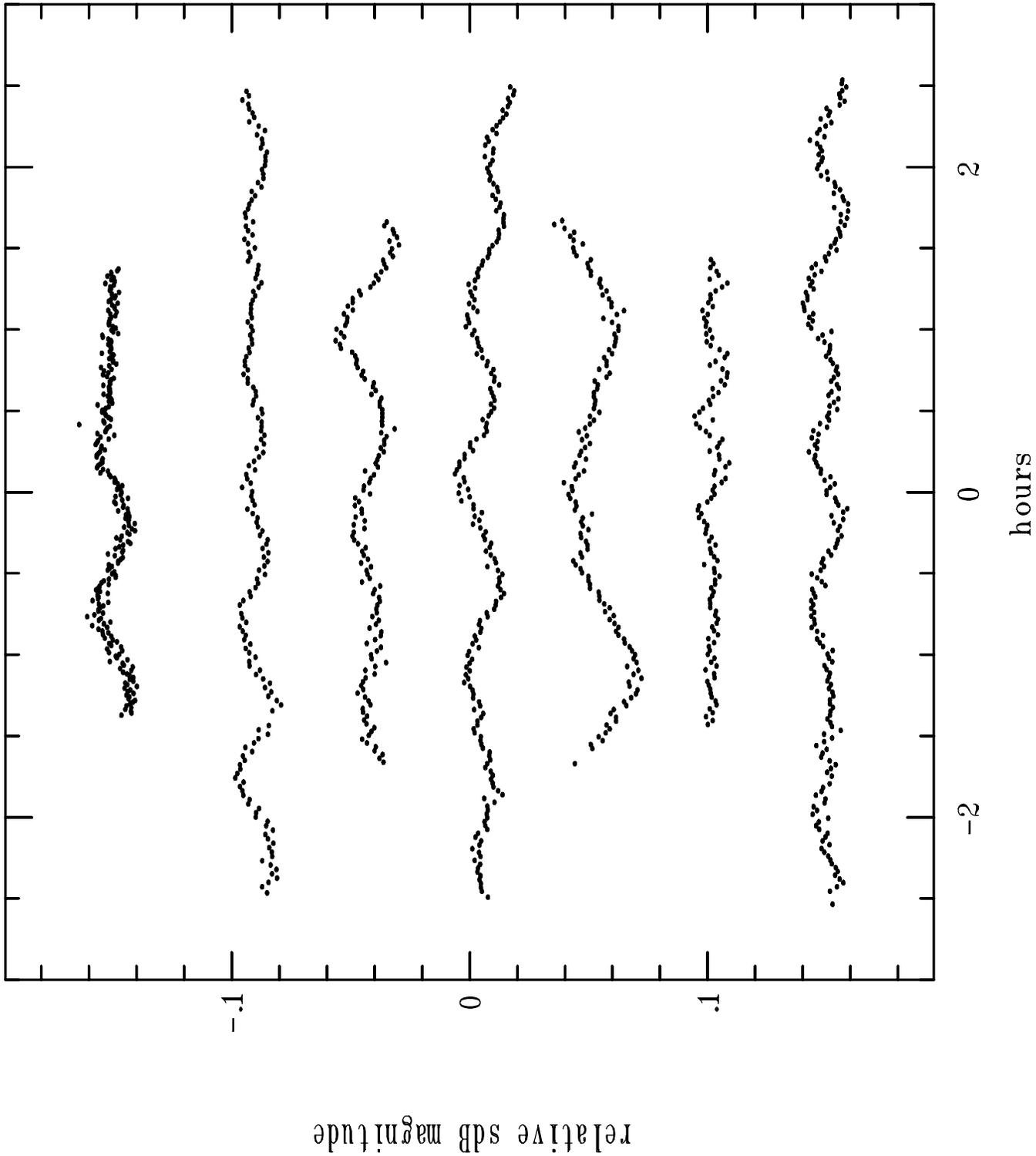
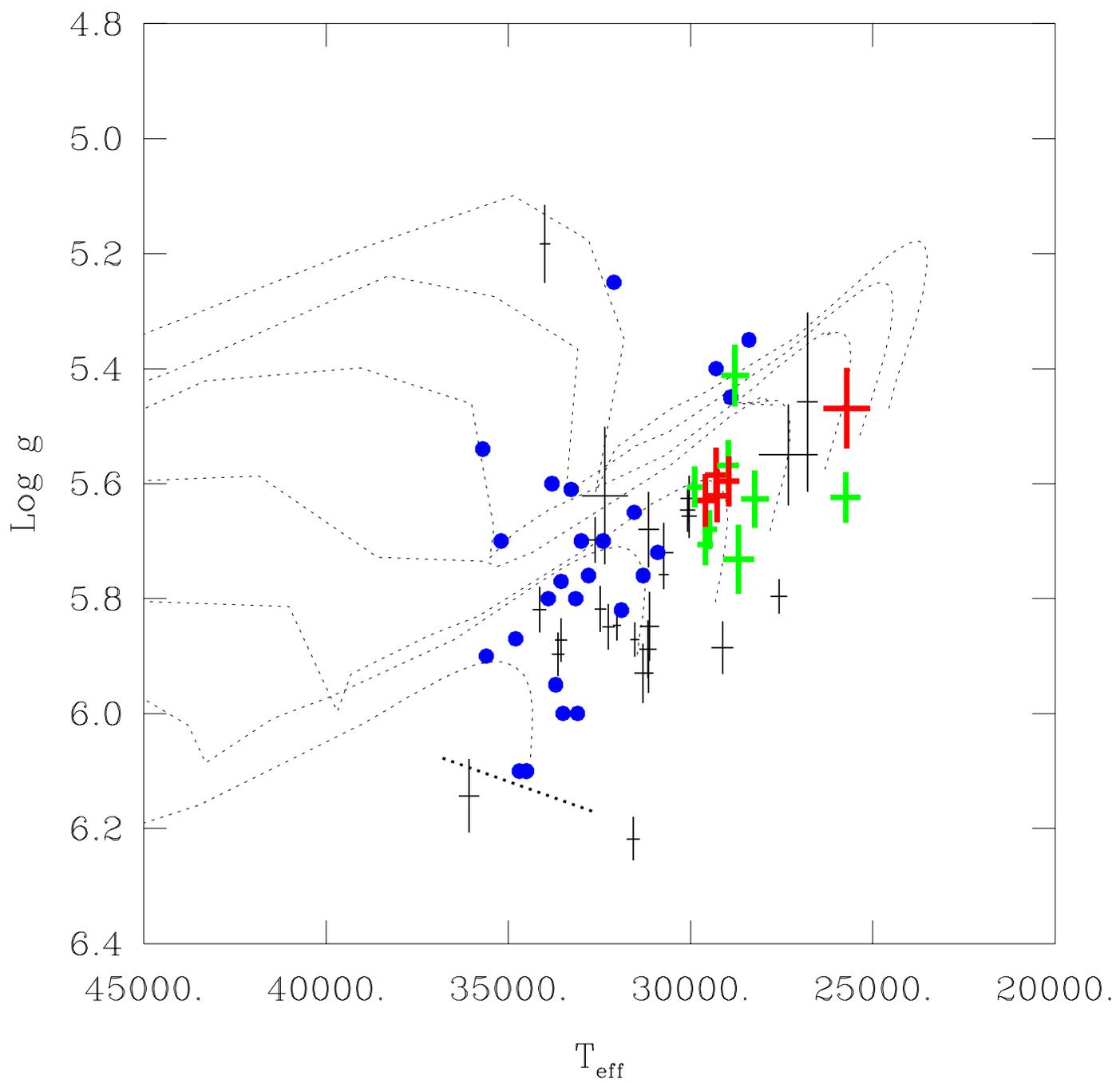
\end{document}